# Evolving Musical Counterpoint

The Chronopoint Musical Evolution System

Jeffrey Power Jacobs
Computer Science Dept.
University of Maryland
College Park, MD, USA
jjacobs3@umd.edu

Dr. James A. Reggia[*]
Computer Science Dept. & UMIACS
University of Maryland
College Park, MD, USA
reggia@cs.umd.edu

*Abstract*—**Musical counterpoint, a musical technique in which two or more independent melodies are played simultaneously with the goal of creating harmony, has been around since the baroque era. However, to our knowledge computational generation of aesthetically pleasing linear counterpoint based on subjective fitness assessment has not been explored by the evolutionary computation community (although generation using objective fitness has been attempted in quite a few cases). The independence of contrapuntal melodies and the subjective nature of musical aesthetics provide an excellent platform for the application of genetic algorithms. In this paper, a genetic algorithm approach to generating contrapuntal melodies is explained, with a description of the various musical heuristics used and of how variable-length chromosome strings are used to avoid generating "jerky" rhythms and melodic phrases, as well as how subjectivity is incorporated into the algorithm's fitness measures. Next, results from empirical testing of the algorithm are presented, with a focus on how a user's musical sophistication influences their experience. Lastly, further musical and compositional applications of the algorithm are discussed along with planned future work on the algorithm.**

*Genetic algorithms; musical counterpoint; subjective fitness; variable-length chromosomes; musical heuristics*

## I. Introduction

Since the baroque era, practitioners of musical counterpoint, from Johann Sebastian Bach and Amadeus Mozart to Arnold Schoenberg and Igor Stravinsky, have experimented with the myriad ways in which two or more melodies can be superimposed on one another to form aesthetically pleasing harmonic lines. Compositional forms such as the fugue and the canon were developed and refined over the past few centuries in order to give composers a structural framework within which to experiment with various contrapuntal techniques. The writing of counterpoint is a fascinating process in that it requires the composer to both express his or her creativity and to adhere to strict music theoretical rules. This fusion of creative energy and structural discipline provides an excellent platform for applying the evolutionary computational approach to the generation of counterpoint.

Several different approaches to computational counterpoint generation have been developed over the past 5 to 10 years. Various genetic algorithms using the deterministic rules of first species counterpoint to objectively evaluate a given melody's fitness have been developed [1, 3]. Machine learning approaches to the problem have also been researched in-depth [2]. In addition, several general composition tools utilizing various evolutionary computation paradigms with objective fitness evaluation have been developed. Such tools include GA's for "melody extension" [4] and for creating melodies from scratch [5]. However, to our knowledge the question of whether subjective or objective fitness evaluation will lead to more aesthetically pleasing counterpoint has yet to be addressed. Therefore, the aim of this work is to elucidate the advantages and disadvantages of using subjective fitness in the context of counterpoint generation.

Although aesthetic value is hard to quantify, there are numerous examples of novel artistic creations generated using evolutionary computation paradigms with subjective fitness evaluation. One very fitting example is GenJam (short for Genetic Jammer), an interactive genetic algorithm for dynamic jazz improvisation [6]. GenJam essentially allows jazz musicians to interact with and "play off of" a computer during live jazz performances. The algorithm has been performing live with its creator Al Biles for about 18 years, and even has a CD available for purchase through Rochester Institute of Technology's campus bookstore.

Attempts to create visual art using EC paradigms have also led to a number of novel innovations. One archetypal example comes from a project undertaken by Witbrock and Neil-Reilly [7], in which users could log on to a website and vote for images, thus essentially having a publicly determined subjective fitness function. This technique generated a large number of very intricate and aesthetically pleasing images. Others have utilized subjectively-guided evolutionary computation to create simple "creatures". Richard Dawkins introduced a simple GA in *The Blind Watchmaker* [8] in which a "biomorphic" algorithm is designed which generates abstract organisms whose shape is determined by a simple "genome", or vector of mathematical values. Building upon Dawkins' example, Sims, Todd, and Latham [9, 10] have implemented fascinating applications of the GA in which evolutionary "creatures" are evolved using hybrid objective-subjective fitness evaluation methods. In these applications, the user can choose based on their aesthetic preferences which individuals survive into the successive generation and which ones die out.

In the context of these prior works in EC applications using subjective fitness evaluation, it is clear that evolving contrapuntal melodies using subjective fitness evaluation has great potential to produce novel and aesthetically pleasing counterpoint. With this in mind, a genetic algorithm for musical counterpoint is studied which can take advantage of a subjective fitness evaluation function to emulate the creative process of contrapuntal composition.

[*]Corresponding Author

## II. METHODS

The "brute force" method of simply generating a random string of notes with random duration values to play simultaneously with a given melody is not an effective way of generating harmonic counterpoint for a number of reasons. The primary issue arises from the existence of musically dissonant intervals. In any given major key, 5 out of the 12 possible notes do not fit into the key, and therefore are considered harmonically dissonant when played simultaneously with the tonic, or root note, of the key (these dissonant intervals are: the minor $2^{nd}$, the minor $3^{rd}$, the 'tritone', the minor $6^{th}$, and the minor $7^{th}$). Therefore, we should develop heuristics in order to avoid or reduce the possibility of generating these notes. In addition to the problem of harmonic dissonance, there is also a less well-defined notion of rhythmic dissonance. Rhythmic dissonance occurs when a melody contains spasmodic or "jerky" changes in its rhythmic contour. For example, a melody containing 4 sixteenth notes in a row followed by a sixteenth note rest followed by a half note and a dotted eighth note rest (Figure 1) would certainly be considered rhythmically dissonant due to its rapid fluctuation between relatively large and relatively short note durations.

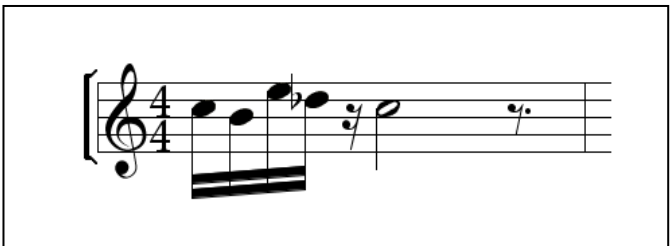

Figure 1.  A rhythmically dissonant phrase

### A. Representation of the Genome

With the above considerations in mind, attempts were made to use a fixed-length chromosome to represent a given melody. However, this method turned out to have several drawbacks. First, the aforementioned rhythmic dissonance problem appeared far too often in melodies coded using fixed-length strings. This dissonance manifested itself in various ways – in some cases jarring rests appeared between long strings of notes, in others spasmodic staccato notes appeared between long rest periods. In either situation, the juxtaposition of the long and the short durations caused far too many unpleasant phrases to be generated.

Given the intractability of the rhythmic dissonance problem with fixed-length strings, the chromosome representation was changed to be a variable-length string. This change allowed for far greater consistency within and between the various phrases of the generated melodies. The major change came in how each individual note event was represented – in the variable-length string, every 10 bits was used to determine one note event. The bits were mapped to note events as follows:

- $B = b_1b_2b_3b_4b_5b_6b_7b_8b_9b_{10}$ is the original bit string.

- $M = b_1b_2b_3b_4b_5b_6$ is used to determine the MIDI value of the note. Only MIDI values between 48 and 83 (inclusive) were used to avoid excessive pitch variation, allowing for 36 unique notes to be represented. So, for example, the MIDI value of $010111_2 \Leftrightarrow 23_{10}$ is 48+23 = 71, or the note B4 (the note a major $7^{th}$ interval away from middle C (C4)).

- The values of M between 0 and 13 (inclusive) and 50 and 63 (inclusive) represent a rest event. This allows for the first 14 and last 14 possible integers to be reserved for special events and the middle 36 integers to be reserved for standard note events.

- $D = b_7b_8b_9b_{10}$ determines the note's duration. The binary encoding of durations is described in Table I.

- Note that one unique integer maps to each type of dotted note, while whole, eighth, and sixteenth notes have 2 integer representations and half and quarter notes have 3 representations. This was done in order to enforce that quarter and half notes are the most frequently occurring notes, followed by whole, eighth, and sixteenth notes, followed by the dotted notes. This strategy is a heuristic decided upon from *a priori* observations of note frequencies in contrapuntal music.

Given the note event representation above, longer strings are constructed by concatenating note event strings. Each measure of the melody can therefore be represented as n concatenated note event strings, where n is an integer ranging from zero to 15. Fifteen was chosen as the maximum number of note events allowed per measure in order to minimize the number of occurrences where the sum of the individual note durations are too high or too low. In order for a measure string to be valid, the sum of its component note event strings' durations must come out to be equal to four quarter notes (all melodies are in 4/4 time, implying that each measure contains four quarter notes). Although this choice of representation presents the issue that some strings produce "invalid" melodies (which do not fit into the allotted 8 measures), it was selected over various other choices due to empirical observations of aesthetic novelty. In other words, although other representations could have allowed for fewer invalid measure strings to be created, the chosen method was found to generate the most "musically interesting" melodies relative to the other methods of representation tested. At the top level of representation, the melody is comprised of eight valid measure strings concatenated together. The optimal melody length was chosen to be eight measures due to empirical observation of how long the melodies could be made before users indicated difficulty in holistically evaluating them.

TABLE I. BINARY ENCODING OF NOTE DURATIONS ($B_7B_8B_9B_{10}$)

| Genotype | Corresponding Phenotype |
|---|---|
| $0000_2$ | Whole note |
| $0001_2$ | Half note |
| $0010_2$ | Quarter note |
| $0011_2$ | Eighth note |
| $0100_2$ | Sixteenth note |
| $0101_2$ | Dotted half note |
| $0110_2$ | Dotted quarter note |
| $0111_2$ | Dotted eighth note |
| $1000_2$ | Dotted sixteenth note |
| $1001_2$ | Whole note |
| $1010_2$ | Half note |
| $1011_2$ | Quarter note |
| $1100_2$ | Eighth note |
| $1101_2$ | Sixteenth note |
| $1110_2$ | Quarter note |
| $1111_2$ | Half note |

*B. The Evolutionary Process*

In evolutionary computation, it is generally desirable to have large populations to provide the substantial diversity needed for effective results. However, when one is determining subjective fitness measures, large populations can soon become very tedious for the evaluator who must listen to each represented melody and rate it relative to others in the population. Accordingly, in evaluating our approach we are using very small populations to obtain results. Out of a wide variety of possible evolutionary processes, two were chosen as the most suitable on the basis of how musically interesting they were and how easily they could be generalized or expanded upon. In addition, the constraint of having such a small population size rendered some possibilities infeasible and thus narrowed the range of effective evolutionary schemes to those which allowed for an appropriate balance of genetic diversity and subjectively "optimal" convergence.

The first process maintains a constant population size (6) by ranking the individuals in each generation by fitness, eliminating the two least fit members, and then using crossover to generate every possible pair-wise combination of the remaining 4 members. Because this method does not allow individuals to survive into the next generation, it generally produced a wide variety of melodic ideas but was somewhat inconsistent. The second process also maintains a constant number of individuals (3) in each generation, but does so through different means. This method ranks the individuals in a given generation by fitness and eliminates the least fit member. It then carries the two fittest members into the next generation, and generates a third member by applying crossover to the two surviving melodies. Because this method allows individuals to survive into the next generation, it was far more consistent than the first method but did not produce as wide a variety of melodic ideas as the first method did. The extremely small population size in this process was clearly a hindrance to melodic diversity, but is somewhat more conducive to allowing the user to exert control over a final "optimal" melody.

Although they differ in how individuals are chosen for each generation, both methods utilize the same crossover and fitness evaluation methods. The crossover function was implemented as follows:

- Let $M_1 = m_{11}m_{12}...m_{1a}$, $M_2 = m_{21}m_{22}...m_{2b}$ be melodies, where $m_{ij}$ represents the $j^{th}$ bit of the $i^{th}$ melody string, and a, b represent the total number of bits in $M_1$ and $M_2$, respectively.

- C, a randomly generated index of $M_1$, is chosen as the crossover point. Two new strings $M_1^*$ and $M_2^*$ are created so that $M_1^* = m_{21}m_{22}...m_{2c}m_{1(c+1)}m_{1(c+2)}...m_{1a'}$ and $M_2^* = m_{11}m_{12}...m_{1c}m_{2(c+1)}m_{2(c+2)}...m_{2b'}$, where a' and b' represent the total number of bits in $M_1^*$ and $M_2^*$, respectively.

- If $M_1^*$ or $M_2^*$ are found to be too long, they are simply heuristically truncated so as to have them fit into 8 measures. If one or both are found to be too short, randomly generated note events in the same key as the melody are appended to the end of the string until the string is of sufficient length. If these truncations or concatenations cause the melody to be too long or too short, the reverse operation is used – truncation is used if concatenation fails, concatenation is used if truncation fails.

Generating the initial population of melodies to be evolved proved to be the toughest problem to tackle from a programming perspective. A way to transform or mutate the notes of the original melody while maintaining melodic and rhythmic consistency with the original melody was needed. To accomplish this, the following algorithm was used:

- Copy the original melody's string M into a new string M*.

- Iterate through the measure strings comprising M*, performing between one and four of the mutation operations described in Table II on each successive pair of note events within the measure strings.

- Performing between one and four of the operations described in Table II for each pair of note events, allows for a total of 15 operations (the sum of the $4^{th}$ row of the binomial coefficients is 16, and we do not allow the possibility of zero operations, giving $2^4 - 1 = 16 - 1 = 15$ total possibilities). Each of the 15 operations is given an equal probability of occurring.

- With a probability of .2, an extra randomly generated note event in the key of the melody is generated and placed in between a pair of note events, thus allowing the mutated melody to have a different number of note events from the original melody. To compensate for the added length, either the prior or the subsequent note is shortened (with probability .5). If it cannot be shortened, it is removed. This essentially corresponds, in evolutionary terms, to a single nucleotide polymorphism (SNP) mutation of the original gene.

TABLE II. MUSICAL MUTATION OPERATORS

| Operator | Description |
|---|---|
| INVERT | Invert the interval between the first and second notes |
| REVERSE | Reverse the order of the notes |
| AUGMENT | Increase the interval by one |
| DIMINISH | Decrease the interval by one |

*C. Experimental Methods*

To perform the actual experiments for empirical testing of the algorithm, a very simple graphical user interface (GUI) was developed to allow users to evolve melodies. A screenshot of the most basic state of the interface is shown in Figure 2. The interface allows the user to listen to a melody however many times they would like and give it a fitness rating from 0 to 100, where 100 represents "most aesthetically pleasing". The user may select "next" to proceed to the next melody, or "previous" to go back and reevaluate a prior melody with the current melody in mind (this allows a user to rate the melodies' fitness scores relative to each other rather than absolutely, a decision which we found worked well when using subjective fitness). In addition, at any point in the evolutionary process the user is given the option of choosing "complete", ending the process and giving them the option of choosing a melody in the current generation as the "final" melody, the subjectively optimal solution in terms of the user's tastes. Otherwise, once all melodies have been assigned a fitness score, the user can move to the next generation by selecting "evolve". They are then brought to the first melody in the new generation, and the process repeats.

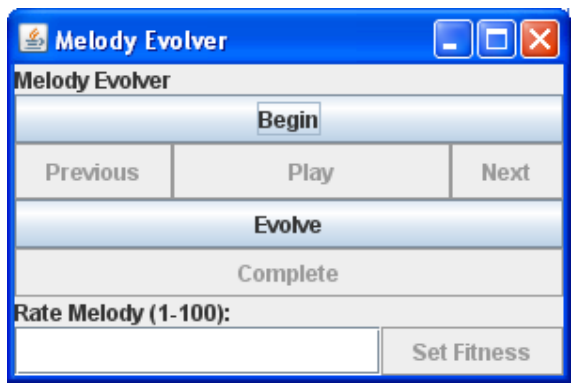

Figure 2. The GUI for the melody evolution process

For the purposes of empirical testing, the application was designed in such a way that users with varying levels of familiarity with music theory would be able to perform the necessary evaluations for the evolutionary processes. This way, the interface could be used for a cross-sectional study to investigate satisfaction with evolved melodies across levels of musical knowledge. The interface was designed as modally as possible to minimize potential ambiguities the as to what the next step of the evolutionary process is.

Before a test session begins, users play the original melody which is to have counterpoint generated for it. In our preliminary tests, the melody chosen was 8 bars of a single voice from Johann Sebastian Bach's "Canon a 2 Quaerendo invenietis" from *The Musical Offering* (Figure 3), colloquially referred to as Bach's "Crab Canon" because in the standard arrangement the two voices play the exact same notes as each other in reverse [11]. This melody was chosen due to the fact that few test subjects are likely to have heard it previously, so they would not be biased by knowing the "correct" counterpoint, and because it was identified by us out of a large pool of Bach melodies as being particularly melodically interesting, thereby minimizing the tedium which would naturally result from other more tonally "straightforward" Bach melodies.

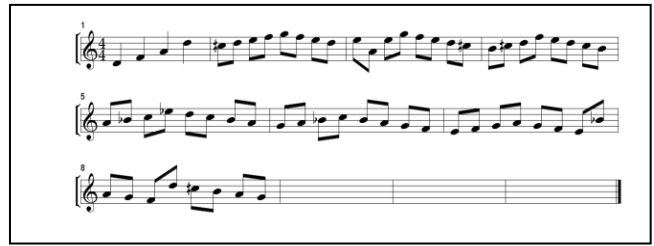

Figure 3. The selection from "Crab Canon" used in the application

Once a user has been played the original melody, they begin the evolution process. If at any time during the process a user decides upon one of the melodies as the "optimal" melody (essentially when the marginal benefit of continuing to a subsequent generation would not be worth the time investment), they can click "complete" and finish the evolutionary process by selecting which of the melodies in the current generation is the most fit.

Our study used a range of human subjects to evaluate the subjective-fitness approach to music evolution. Users were selected for the study across a wide range of pertinent attributes such as musical knowledge, cultural musical tradition (diatonic vs. chromatic, microtonal vs. equal tempered, etc.), and relative aesthetic preference for harmony versus dissonance.

III. RESULTS

In this section, we describe three sets of experimental trials we performed using our algorithm: one with a beginning-level user with no prior musical experience, one with a novice-level user with some basic formal musical knowledge, and one with an advanced-level user with significant prior formal musical training. The subjects evolved as many generations as they needed until they indicated satisfaction with one of the melodies and chose it as their "final" melody. Each generation consisted of six contrapuntal melodies, played along with the original base melody (the "Crab Canon").

*A. The Beginning-Level User*

This user had no prior experience in formal music theory. Our hope was that this user would be able to develop interesting and aesthetically pleasing contrapuntal melodies with the assistance of our application. Though this user stated that they had fun composing with the application, they also said that it was somewhat difficult for them to rate the melodies relative to one another, as a lot of the melodies in a given generation sounded too similar to the user for them to pick out which melodies they liked and disliked. Overall, however, the user was extremely satisfied with the final melody, formed over 15 generations, and gave it a score of 91. An illustration of the final melody evolved by this user is given in Figure 4.

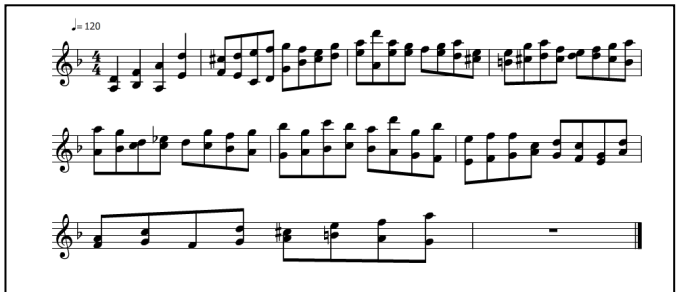

Figure 4. Counterpoint generated during the beginning-level user trial

*B. The Novice-Level User*

This user had some prior experience listening to and analyzing musical counterpoint, but had never attempted composing an original contrapuntal piece. Our hope was that this user would have the opportunity to gain experience in contrapuntal composition with the assistance of our application. Of all three trial users, this user seemed to get the most out of our application. They stated that they were very quickly able to evolve a melody that they were quite satisfied with. The user finished the evolution process after nine generations, and gave the final melody a rating of 84. The generated melody is illustrated in Figure 5.

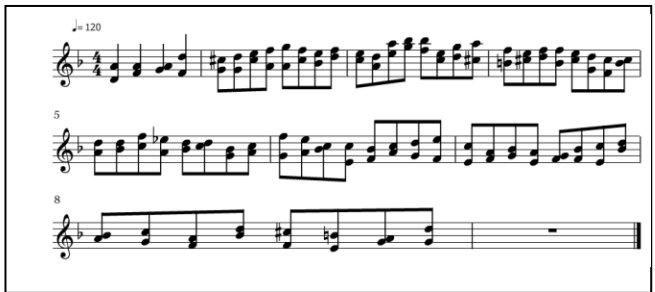

Figure 5. Counterpoint generated during the novice-level user trial

*C. The Advanced-Level User*

This user had extensive prior experience listening to, analyzing, and composing contrapuntal pieces. Our hope was that the application would allow this user to explore various different melodies superimposed over a melody they were already quite familiar with – the "Crab Canon". The user's general feelings towards the application were that it was simple to use but the time and effort necessary to evolve a melody from scratch was a bit cumbersome. The user finalized the evolution process after 11 generations, and gave a score of 67 to the final melody. An illustration of the melody is given in Figure 6.

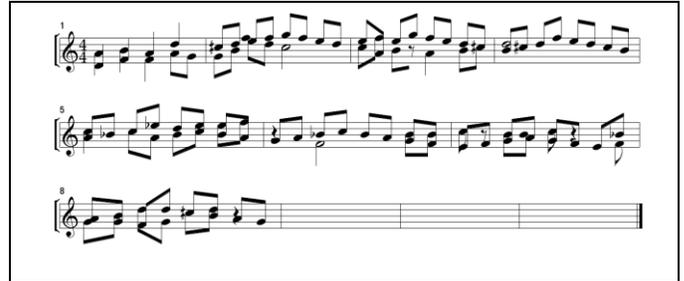

Figure 6. Counterpoint generated during the advanced-level user trial

An auxiliary question which we investigated throughout our trials is what type of interface was more helpful to the user: one in which the user can listen to the melodies and view them in standard musical notation or one which only allows the user to listen to the melodies. Allowing the user to see visual representations of the melodies seemed to cause the more musically knowledgeable user to be biased in the sense that melodies with more accidentals (sharp, flat, or natural symbols adjacent to a note indicating that the note is modified from its natural scale tone) seemed to negatively impact their evaluations of the melodies, which was also indicated to us during discussions with them. Once the visual display was removed from the interface, the subject stated that they were better able to judge the melodies based on their aural qualities alone. The user with basic musical knowledge surveyed after their trials did not indicate the presence of a bias regardless of whether the visual representations of the melodies were displayed or not. This is a topic which we plan to investigate further in later trials.

IV. DISCUSSION

Despite being in its early stages, the algorithmic approach to generating musical counterpoint with subjective fitness evaluation outlined above has already proven to be a novel solution to the problem of computationally generating aesthetically pleasing musical counterpoint. The emulation of the extremely complex yet deterministically constrained process of contrapuntal composition by a genetic algorithm is a lofty goal, but one which we believe will have great significance to both musicians and computer scientists once it is achieved. Although previous work using genetic algorithms for counterpoint generation with objective fitness evaluation has elucidated the many possible ways the problem can be approached from a purely objective computational standpoint, we believe that the intersection of both objective computation and subjective aesthetic evaluation is where an elegant solution to the problem can be found. Through further expansion of the basic algorithm presented in this paper, we believe that it will come to fruition as an invaluable tool for composers and an extremely successful application of the genetic algorithm paradigm in the eyes of evolutionary art practitioners. The application can also be an entertaining way for interested students to learn the basics of genetic algorithms. Instructors in basic artificial intelligence-related courses or workshops could use the application as a hands-on introduction to the various aspects which comprise a GA.

At this point in time, there are significant limitations of the algorithm which must be eliminated before it can be expected to achieve peak performance. A brief list of the most significant extensions that we plan to make is as follows:

- *Ability to evolve non-4/4 time signatures*: Although 4/4 is the predominant time signature in Western classical music, a counterpoint generation application would not be complete without the ability to create contrapuntal melodies for waltzes in 3/4 as well as 2/4 marches and common complex time signatures such as 5/4 and 7/4. In our planned extension, the user would have the option of specifying a time signature of their choice prior to beginning the evolutionary process.

- *Support for chromatic passing tones*: While much of Western contrapuntal music is dominated by diatonic (7-note, octave-repeating) scales, there exist many examples of contrapuntal melodies utilizing chromatic passing tones (essentially non-harmonic notes placed between two harmonic notes to embellish the transition between the two notes) [12]. Currently, the algorithm only produces note events corresponding to notes in the key of the composition.

- *Support for melodies containing modulations or tonicizations*: Currently, the algorithm keeps a static record of the melody's key, and only generates note events corresponding to notes in this key. This poses a significant problem, however, when trying to run the algorithm with melodies containing modulations (key changes) or tonicizations (temporary treatment of non-tonic notes as tonic notes). Incompatibility with modulation should not be an intractable problem, however, as its solution would simply require the mutation algorithm to have a dynamic record of the current key. Tonicization is somewhat more difficult to account for as there is generally very little information given as to the new temporary tonic or root note when a tonicization occurs [13].

- *Objective Evolution*: To facilitate further investigation as to the efficacy of our trials, we will need to implement an objective evolution scheme for comparison purposes. Users should be able to evolve and rate one subjectively generated melody, then rate another objectively generated melody derived from the same base melody so that the marginal difference can be quantified. Existing objective approaches to contrapuntal evolution [1, 3] can be utilized towards this end.

- *Improved musical heuristics*: Although we attempted to design the musical heuristics utilized by the algorithm through careful empirical analysis and logically sound *a priori* reasoning, it is inevitable that problems with these heuristics will become clear over time. Therefore, as we test the algorithm over a wide variety of melodies the accurate emulation of human-created counterpoint will be the primary focus of future modifications to the algorithm's heuristics.

ACKNOWLEDGMENT

We thank Jefferson Teng, Josh Fann, Chris Parker, and Kelly Jackson for their consistent encouragement and valued input throughout the research and testing process.